\RequirePackage{amsmath}

\documentclass[runningheads]{llncs}
\usepackage[T1]{fontenc}
\usepackage{graphicx}
\usepackage{booktabs}
\usepackage{multirow}
\usepackage{nicematrix,tikz}
\usepackage{pbox}
\usepackage{subcaption}
\usepackage{csquotes}
\usepackage{newclude}
\usepackage{hyperref}
\usepackage{color}

\begin{document}
\title{Interpretability of Uncertainty: Exploring Cortical Lesion Segmentation in Multiple Sclerosis}

\author{
Nataliia Molchanova\inst{1,2,3}\orcidID{0000-0002-7211-8863} \and 
Alessandro Cagol\inst{4} \and 
Pedro M. Gordaliza\inst{1,3} \and 
Mario Ocampo-Pineda\inst{4} \and
Po-Jui Lu\inst{4} \and
Matthias Weigel\inst{4} \and
Xinjie Chen\inst{4} \and
Adrien Depeursinge\inst{1,2} \and 
Cristina Granziera\inst{4} \and 
Henning Müller\inst{2,6} \and 
Meritxell Bach Cuadra\inst{1,3}
}
\authorrunning{Molchanova et al.}
\titlerunning{Interpretability of Instance-Wise Uncertainty in Multiple Sclerosis}
\institute{
University of Lausanne and Lausanne University Hospital, Switzerland
\and
University of Applied Sciences Western Switzerland (HES-SO), Switzerland
\and
CIBM Center for Biomedical Imaging, Switzerland
\and
University Hospital and University Basel, Switzerland
\and
University of Genova, Italy \and University of Geneva, Switzerland
}

\maketitle              
\begin{abstract}

    Uncertainty quantification (UQ) has become critical for evaluating the reliability of artificial intelligence systems, especially in medical image segmentation. 
    This study addresses the interpretability of instance-wise uncertainty values in deep learning models for focal lesion segmentation in magnetic resonance imaging, specifically cortical lesion (CL) segmentation in multiple sclerosis.
    CL segmentation presents several challenges, including the complexity of manual segmentation, high variability in annotation, data scarcity, and class imbalance, all of which contribute to aleatoric and epistemic uncertainty.  
    We explore how UQ can be used not only to assess prediction reliability but also to provide insights into model behavior, detect biases, and verify the accuracy of UQ methods. Our research demonstrates the potential of instance-wise uncertainty values to offer post hoc global model explanations, serving as a sanity check for the model. The implementation is available at \url{https://github.com/NataliiaMolch/interpret-lesion-unc}.

\end{abstract}
\section{Introduction}
Uncertainty quantification (UQ) is gaining popularity within the field of medical image segmentation as a means to assess the reliability of artificial intelligence systems by representing the "degree of untrustworthiness" of their predictions~\cite{osti_1561669, lambertreview, uncs_survey}. Higher uncertainty in a prediction indicates an increased likelihood of an erroneous prediction. Consequently, a common UQ evaluation practice involves assessing the correspondence between uncertainty and error, using methods such as uncertainty calibration measures~\cite{10.3389/fnins.2020.00282}, accuracy-confidence curves~\cite{DojatMICCAI}, or error retention curves~\cite{molchanova_isbi}. 
Uncertainty can thus be effectively used for assessing the quality of predictions at inference time without the need for ground truth~\cite{10.3389/fnins.2020.00282,DojatMICCAI}. 
UQ has also been applied in other downstream tasks, including active learning and domain adaptation, among others~\cite{lambertreview,uncs_survey}.

While UQ is actively used for various downstream tasks, little has been done to analyze and interpret the uncertainty values themselves~\cite{10.1145/3461702.3462571}. 
Additional analyses providing insights into uncertainty would be highly valuable for:
i) detecting biases in deep learning (DL) model behavior;
ii) performing a sanity check of the UQ methods themselves;
iii) extracting information captured by uncertainty beyond errors.

In this work, we explore the interpretability of instance-wise uncertainty values in DL segmentation within the context of focal lesion segmentation from magnetic resonance imaging (MRI). 
Specifically, we focus on UQ in CL segmentation, a key task for differential diagnosis and prognosis in multiple sclerosis (MS) ~\cite{mac}.

\begin{figure}
    \centering
    \includegraphics[width=\textwidth]{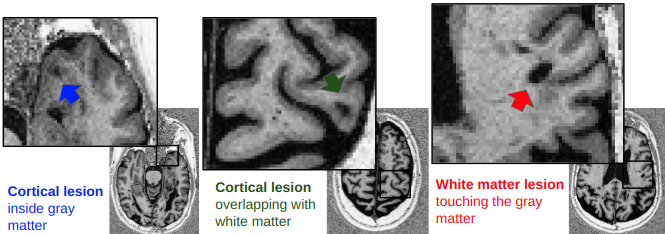}
    \caption{Examples of several types of MS lesions visible on MP2RAGE scans, appearing as hypointense regions.}
    \label{fig:intro}
\end{figure}

Automating CL segmentation is complicated by poor data quality. The ground-truth annotations are subject to errors and high intra- and inter-rater variability due to small lesion sizes and confusion with white matter lesions adjacent to the cortex (see examples in Figure ~\ref{fig:intro}). The data is sparse and limited to private cohorts with varying data characteristics. Additionally, significant class imbalance affects machine learning solutions. These factors hinder the development of DL models, which face two main sources of uncertainty: data noise (aleatoric uncertainty) and training data scarcity and/or domain shifts (epistemic uncertainty)~\cite{uncs_survey}. In this context, the interpretability of uncertainty provides post hoc global model explanations, serving as a sanity check for both the model and the uncertainty values themselves.

\section{Materials and methods}

\subsection{Data}
We use a dataset obtained at University of Basel, Switzerland and previously explored in other studies~\cite{LAROSA2020102335, molchanova_preprint}. 
Our cohort includes 117 patients diagnosed with MS~\cite{mac} at different stages of the disease: relapsing-remitting (62), primary-progressive (39), and secondary-progressive (16). The male-to-female ratio is 0.77; the median Q2 (Q1-Q3) age: 51 (40-58) years, disease duration: 8.7 (1.7-18.2) years, and the expanded disability status scale: 3 (2-6).
All brain MRI scans are obtained with a standardized acquisition protocol on a 3 Tesla whole-body MR system (Magnetom Prisma, Siemens Healthineers), using a 64-channel phased-array head and neck coil for radiofrequency reception. The protocol includes the acquisition of 3D magnetization-prepared 2 rapid gradient-echo (MP2RAGE) images (TR/TI1/TI2=5000/700/2500 ms; resolution = $1 \times 1 \times 1$ mm$^3$). 
Brains are extracted using the HD-BET masks from FLAIR scans registered to the MP2RAGE space.
The annotations are formed by consensus of two medical doctors with expertise in neuroimaging and include intracortical lesions (in gray matter) and leukocortical lesions (intersecting with white matter). 
The dataset was split into training, validation, and test sets in the proportion of 79:8:30 patients, corresponding to 859:69:302 CLs.

\subsection{Cortical lesion segmentation model}
There exist few machine learning models that tackle CL segmentation~\cite{LAROSA2020102335}, and they do so jointly with the identification of white matter lesions (WMLs) within the same \textit{lesion} class. 
Given the clinical importance of the CL biomarker for differential diagnosis~\cite{mac}, we propose a model dedicated solely to CL segmentation.
We adopted a 3D shallow U-Net architecture based on a baseline model from the UQ WMLs segmentation challenge (Shifts Challenge~\cite{shifts20}). 
We modified this model to improve segmentation performance and address specific challenges of CLs, such as small sizes and data sparsity. 
Specifically, we adjusted data augmentation strategy to minimize the distortion of small lesions; replaced the Dice focal loss with focal loss with appropriate weighting; adopted a more effective training strategy (warm-up epochs, a learning rate plateau scheduler, and early stopping). 
The probability thresholds for all models were chosen on the 5-fold cross validation (CV) and set to 0.55. 

\subsection{Uncertainty quantification}
We focus on deep ensemble (DE) and Monte Carlo Dropout (MCDP), two UQ methods widely explored for medical imaging tasks in general and in MS lesion segmentation specifically~\cite{lambertreview, DojatMICCAI, TAlber}. 
DE involves training several similar networks with varied random initialization seeds, which affect random augmentation, weight initialization, training example sampling, and stochastic optimization. This approach allows for obtaining different samples from the posterior distribution of model parameters. 
MCDP was initially designed as a way to perform variational inference by placing dropout layers between the neural network layers and treating different dropout masks as random variables, inducing a distribution over the model's weights. 
DE has been shown to provide better UQ results in terms of the relationship with errors~\cite{uncs_survey}.

For both methods, the final prediction is formed as the mean average across $M$ sampled predictions, and uncertainty is quantified by assessing the spread of these predictions. 
Commonly, for classification tasks, information-theory-based measures like entropy and mutual information are used to quantify uncertainty. For segmentation, classification measures can be used at the pixel/voxel scale. Additionally, it is possible to quantify uncertainty associated with a set of voxel predictions, such as for segmented instances/structures or even for the entire prediction. 

In this work, we aim to explore instance-scale uncertainty and its interpretability. Thus, we focus on computing uncertainty for each predicted lesion. 
Several approaches have been proposed: aggregation (\textit{e.g.} averaging) of voxel-wise uncertainty values~\cite{10.3389/fnins.2020.00282, TAlber}, graph neural networks~\cite{DojatMICCAI}, and disagreement in structural predictions~\cite{molchanova_isbi}. 
We chose a UQ metric based on structural disagreement, which has been shown to better capture lesion detection errors compared to aggregation-based measures~\cite{molchanova_isbi, molchanova_preprint}. 

In a nutshell, given a predicted lesion $L$ and corresponding lesions from the $m^{th}$ sampled prediction $L^m, m=0, 1, 2, ..., M-1$, the lesion structural uncertainty is defined as:
\begin{equation*}
    LSU=1 - \frac{1}{M} \sum\limits_{m=0}^{M-1} IoU(L, L^m).
    \label{eq:lsu}
\end{equation*}
The corresponding lesions are defined as the ones with maximum intersection over union (IoU).

The number of ensemble members and MCDP samples is chosen to be $M=10$, selected from 3, 5, 7, 10, based on joint uncertainty-robustness assessment (proposed in \cite{molchanova_preprint}) for the DE model. We use a dropout probability of 0.1 for MCDP, chosen among 0.01, 0.05, 0.1, 0.15, 0.2, and 0.25 as the maximum dropout probability that does not yield a significant drop in model prediction quality (quality-diversity trade-off).

\subsection{Interpretability analysis for lesion uncertainties}
The proposed interpretability analysis for lesion uncertainties consists of explaining lesion uncertainty in terms of lesion-related features.

\textbf{Lesion features - }Each predicted lesion is characterized by the following features: i) intensity, ii) texture, iii) shape, iv) location in the brain, and v) segmentation quality. 
Radiomic features from the PyRadiomics Python Library (v3.1.0) are used to characterize the intensity, texture, and shape (i-iii). 
The location (iv) in the brain is characterized using the MNI atlas separated into right (R) and left (L) hemispheres~\cite{grabner-2006}. 
The location feature is computed as the distance between the center of a predicted lesion and the center of the brain structure it belongs to; features corresponding to the rest of the brain structures are zeroed. The belonging of a CL to the MNI brain structure is decided by the maximum overlap. 
The lesion segmentation quality (v) is evaluated using the adjusted intersection over union (IoU$_{adj}$) measure~\cite{9206659}, which is similar to IoU but corrected for overlaps explained by other predicted instances. 
Recursive features elimination with the decision trees is used for feature selection. 

\textbf{Uncertainty regression model - }To explain the lesion-scale uncertainty in terms of the aforementioned features, we use a linear regression model, ElasticNet, which combine $L_1$- and $L_2$-regularization.
Given that all the features are normalized prior to model fitting, the coefficients of the linear model are interpreted as \textit{feature importance}. 
Model selection and feature importance are computed 10 times with different random seeds to assess the standard error. 
A five-fold CV procedure is used to tune the parameters of the pipeline comprising feature selection and ElasticNet model (fraction of selected features, $L_1/L_2$ ratio, and intercept parameters) by optimizing the coefficient of determination ($R^2$). CV is performed on the training set with an evaluation on the test set.

\section{Results}
The CL detection of the DL models reaches a 0.55 F1-score (computed as in~\cite{DojatMICCAI}), which is a good performance considering the high inter-rater variability of CL manual segmentation (Cohen $\kappa \in [0.4, 0.6]$ depending on the study)~\cite{madsen-2021}.

The regression quality ($R^2$) of the uncertainty interpretability model is shown in Table~\ref{tab:elasticnet} for different sets of features: only IoU$_{adj}$ (v), only radiomics and location (i-iv), and all together (i-v). For both DE and MCDP uncertainty, the $R^2$ improves when using all features compared to other settings. For MCDP, less variation in uncertainty can be predicted (worse quality fit) compared to DE, regardless of the features used. For both DE and MCDP uncertainty, some variability in uncertainty is unexplained in terms of chosen features.

\begin{figure}[ht!]
\centering
    \begin{subfigure}{\textwidth}
        \includegraphics[width=.9\textwidth]{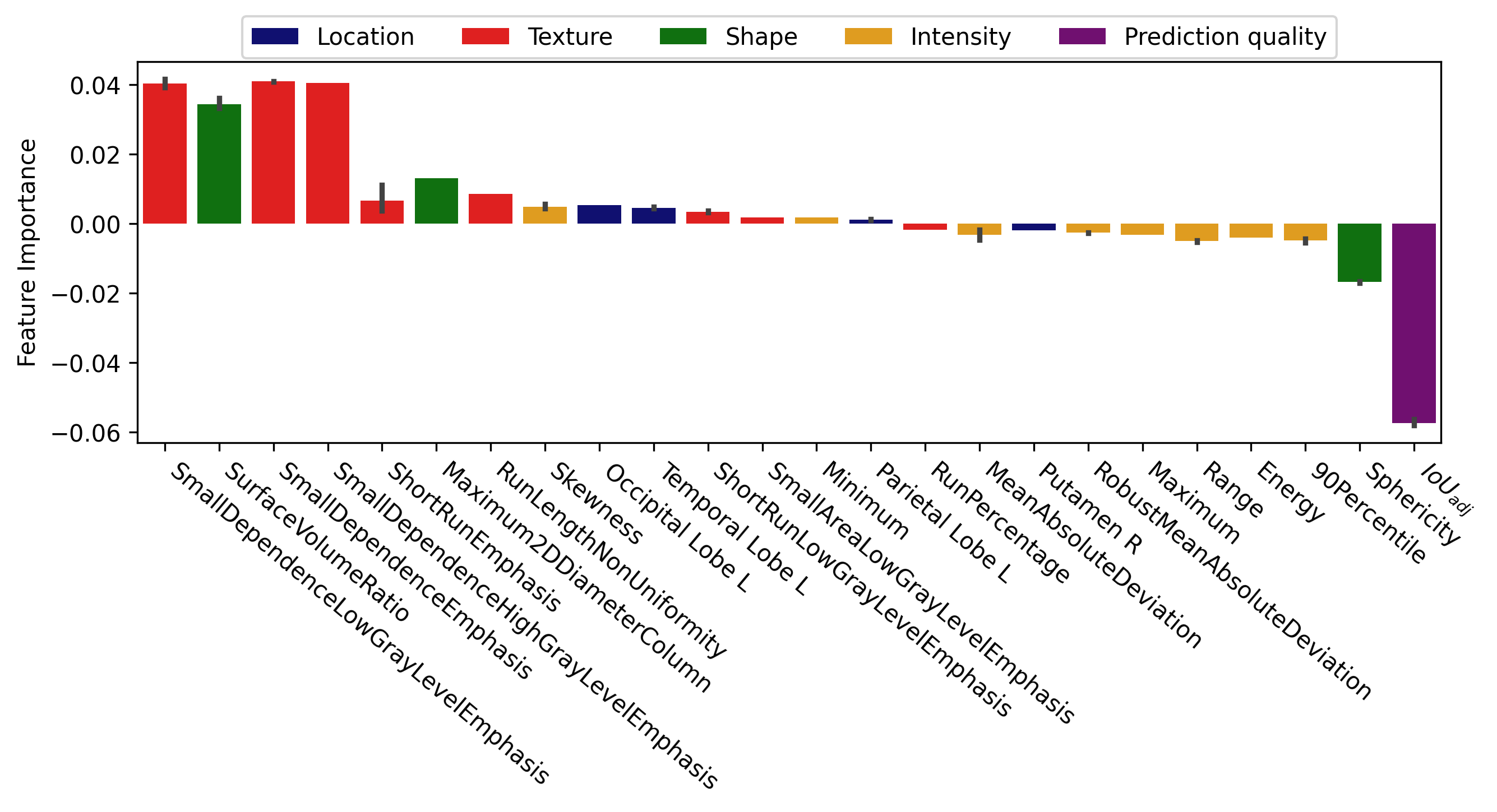}
        \vspace{-.4cm}
        \caption{DE}
        \label{fig:de}
    \end{subfigure}
    ~
    \begin{subfigure}{\textwidth}
        \includegraphics[width=.99\textwidth]{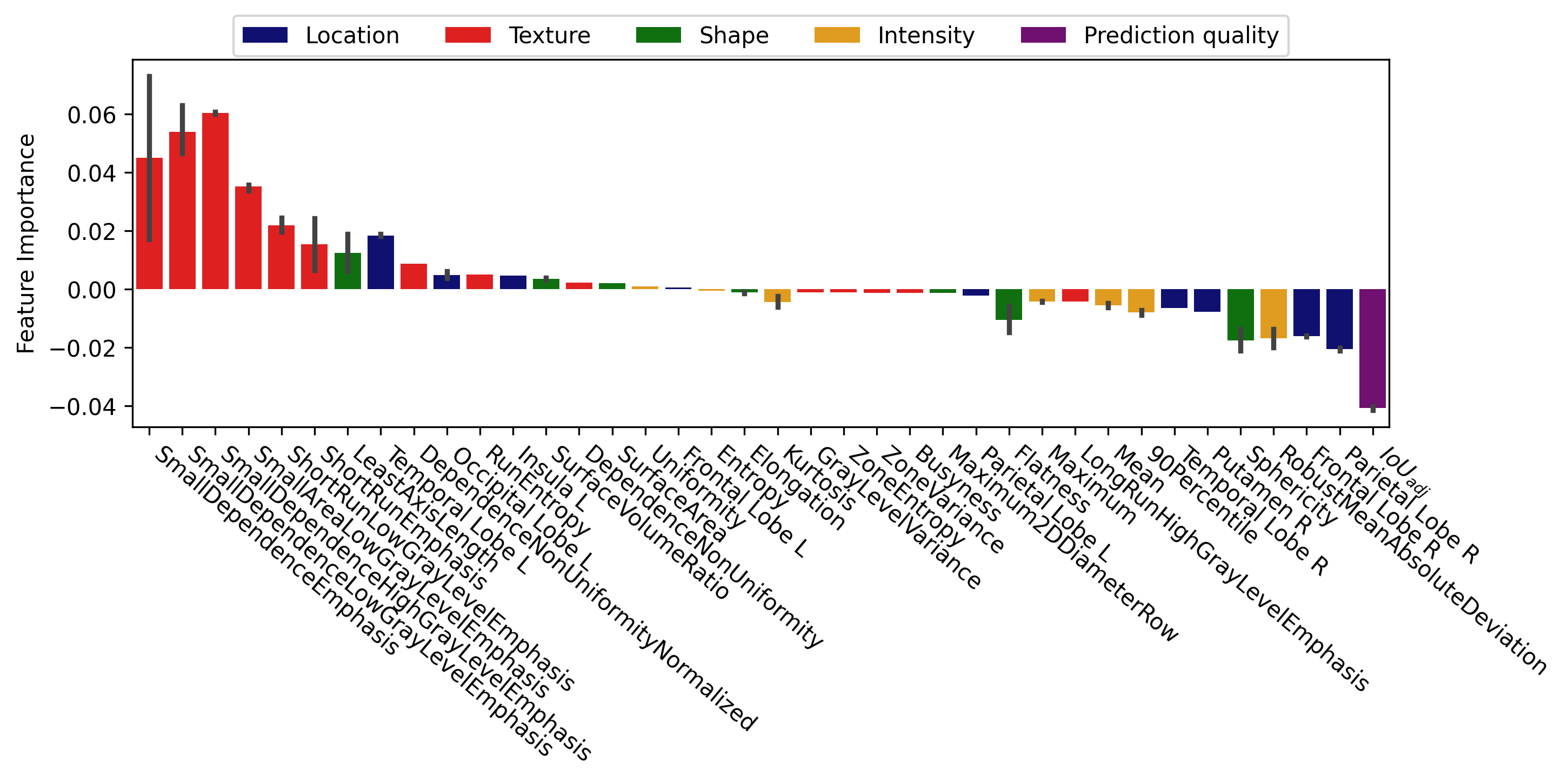}
        \vspace{-.5cm}
        \caption{MCDP}
        \label{fig:mcdp}
    \end{subfigure}
    \caption{
Coefficients of a linear regression model for  explaining lesion uncertainty, averaged over 10 model fits, with standard error. Positive values indicate higher uncertainty, negative - lower.}
    \label{fig:results1}
\end{figure}

The relative feature importance of different features is shown in Figure~\ref{fig:results1}. Overall, the model explaining the MCDP uncertainty selected more features than DE. 
For DE uncertainty, the prediction quality explains most of the uncertainty, indicating a strong relationship between uncertainty and error. For MCDP uncertainty, the texture features have more importance than prediction quality (greater values of linear model coefficients). 
Features selected for both UQ methods resemble, including 
a strong positive relationship of lesion uncertainty with the presence of texture (SmallDependence...Emphasis, ShortRun...Emphasis); 
a positive relationship with lesion elongated shape and negative with sphericity (SufaceVolumeRatio, Maximum2DDiameterColumn, Shpericity, Flatness) and lesion small sizes (SufaceVolumeRatio, LeastAxisLength); 
a negative relationship with features indicating high intensities in a hypointense CL (90Percentile, Energy, Maximum); 
a positive relationship with left brain lobes locations (Temporal and Occipital L). 
\begin{table}[h!]
  \centering
  \small
    \begin{tabular}{lcccccc}
        \toprule
         & \multicolumn{3}{c}{CV} & \multicolumn{3}{c}{Test}  \\
         & Only IoU$_{adj}$ & No IoU$_{adj}$ & All &  Only IoU$_{adj}$ & No IoU$_{adj}$ & All \\
         \midrule
        DE & $0.520_{\pm 0.006}$ & $0.598_{\pm 0.004}$ & 
 $0.661_{\pm 0.004}$ & $0.431_{\pm 0.001}$  & $0.512_{\pm 0.002}$ & $0.632_{\pm 0.004}$ \\
        MCDP &  $0.393_{\pm 0.006}$ & $0.589_{\pm 0.014}$ & $0.604_{\pm 0.013}$ & $0.261_{\pm 0.003}$ & $0.425_{\pm 0.013}$ & $0.494_{\pm 0.004}$\\
        \bottomrule
    \end{tabular}
    \caption{
   $R^2 (\uparrow)$ of a linear model explaining lesion uncertainty computed on CV (averaged over 10 model fits) and on the test set (averaged across patients). Features used to fit the linear models: only prediction quality (Only IoU$_{adj}$); all features except for the prediction quality (No IoU$_{adj}$); all features (All).
    }
    \label{tab:elasticnet}
\end{table}

\begin{figure}[t]
    \centering
    \includegraphics[width=\textwidth]{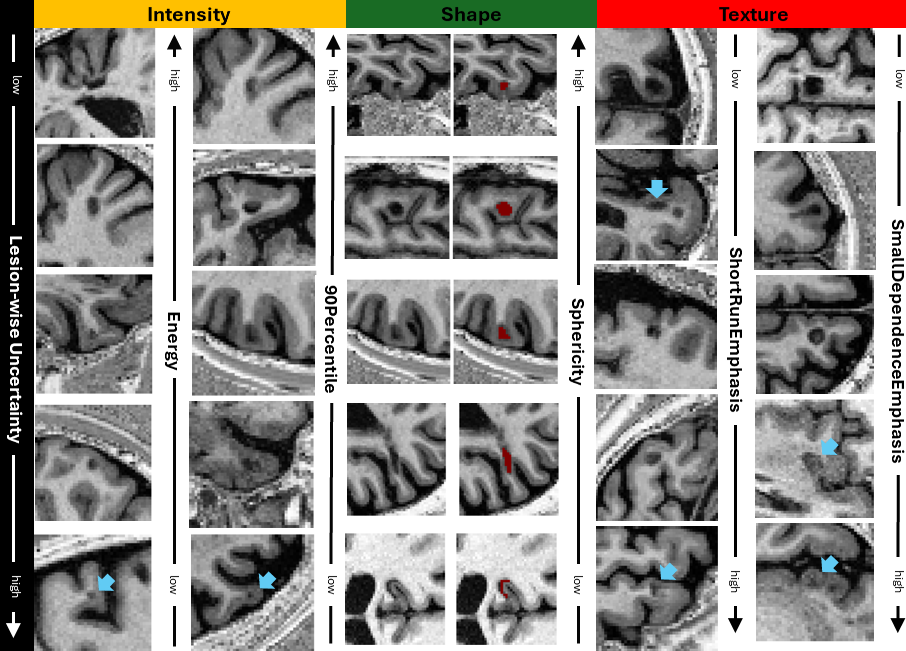}
    \caption{Qualitative results: visualization of true positive CLs with different important feature activation. CLs are T1-hypointense regions in the center of each image, marked with blue errors in doubtful cases. Long black arrows indicate the direction of features growth (from low to high feature values), white arrow indicates the direction of uncertainty growth.}
    \label{fig:results3}
\end{figure}

\section{Discussion}
\textbf{Explained uncertainty - }The \textit{prediction quality} explains a significant portion of the variability in lesion uncertainty. 
However, the inclusion of lesion-related features provides additional information, helping to predict even more variability. This could be due to either the effect of errors in the ground truth or the fact that uncertainty carries more information than just prediction quality.

The \textit{texture features} selected as important and associated with higher uncertainty are derived from gray level dependence matrices (GLDM) and gray level run length matrices (GLRLM) methods of quantifying texture. 
Higher values of GLDM SmallDependence...Emphasis and GLRLM RunLengthNonUniformity indicate the presence of textural patterns and less homogeneous textures. Some texture features are typically higher for smaller regions of interest. 
Both factors complicate the visual identification of lesions (see Figure~\ref{fig:results3}).

The selected \textit{shape features} describe two different lesion profiles. Less spherical, more spiculated, elongated (low Sphericity, high SurfaceVolumeRatio, and high Maximum2DDiameter) lesion shapes are associated with higher uncertainty. 
SurfaceVolumeRatio also tends to be high for small lesions, which are harder to detect visually during the annotation process. 
The elongation and irregularity can be related to the burden of delineation due to the greater surface.

The \textit{location features} are prioritized by the model explaining the MCDP uncertainty, rather than DE uncertainty. Both uncertainty types have an association with lesions located on the periphery of the left occipital and temporal lobes. For MCDP, the proximity to the centers of the right frontal, parietal, and temporal lobes and putamen correlated with more confidence in predictions.

The \textit{intensity features} indicating the presence of high-intensity voxels in a lesion (90Percentile and Energy) are associated with lower uncertainty.

\textbf{Clinical perspective - }Clinical feedback on the visualized lesions (Figure \ref{fig:results3}) was provided by an expert neurologist. 
Lesions associated with higher uncertainty (smaller sizes, heterogeneous intensity with texture blending into the surrounding area) were candidates for lower confidence of raters and are more likely to be overlooked during the manual annotation process. 
Among the lesions with low sphericity feature, many likely represented subpial lesions. These lesions have a distinct pathogenesis and shape, are less evident, and thus less prevalent on 3T MRI. 
Subpial and inhomogeneously-shaped lesions, less common in MS, are likely under-represented in the training data.

The association between higher intensities and lower uncertainty can be due to the neighboring white matter, which has higher intensity on MP2RAGE. CLs appear hypointense on MP2RAGE, thus neighboring white matter increases contrast and visibility of lesions. 
Additionally, from a radiological perspective, higher intensities within a lesion-gray-matter overlap help confirm that the lesion is a true CL rather than a pure white matter lesion.

The characterization of CL locations is still poorly understood, hence the right and left hemispheric differences shown in our analysis are difficult to justify. 
Lesions on the periphery of the temporal and occipital lobes being more uncertain might indicate proximity to the cerebellum and brain stem, hence less common locations with worse lesion-surrounding contrast. These lesions are also likely under-represented in the training data.

\textbf{Unexplained uncertainty - } The $R^2$ reaches 0.661, meaning that some variability in lesion uncertainty is left unexplained.
There are several potential reasons.
First, the fact of using a linear model to explain the non-linear relationships between uncertainty and features. 
We repeated the analysis with a non-linear random forest model, but this did not significantly improve the regression quality or alter the importance of the features. 
Second, the lack of informative features. We use lesion features that encompass characteristics of different natures and approximate the clinical perspective. However, more features could be added to describe the lesion surroundings or to introduce additional clinical context, such as patient information or MS lesion subtypes. 
Finally, interpretability also suffers from the noise in the UQ, related to the quality of the UQ itself. This may explain the higher $R^2$ of the DE uncertainty model compared to MCDP, serving as a sanity check for the UQ.

\section{Conclusions}

We explore the interpretability of instance-wise uncertainty within the context of cortical lesion DL-based segmentation in MS. 
To explain lesion uncertainty, we use an ElasticNet regression model with radiomics, location, and ground-truth overlap features. 
Our results demonstrate how explaining the predicted lesion uncertainty in terms of lesion-related features can:
i) detect model biases towards over- or underperforming on specific types of CLs;
ii) validate the sanity of UQ by assessing the unexplained variability in uncertainty;
iii) facilitate the visual qualitative assessment of the model, helping to select informative lesion examples.

We observe that lesion-scale uncertainty cannot be solely explained by the quality of CL segmentation. 
Given the clinical feedback, we conclude that the features associated with higher uncertainty describe the lesions that are harder to annotate for radiologists and are often less common. 
Thus, we hypothesize that uncertainty would be better explained by the inter-rater variability or rater confidence rather than the ground-truth overlap, as it is dominated by aleatoric or label uncertainty~\cite{lambertreview}.
Our future work should verify those hypotheses.

It is worth mentioning that the proposed analysis show correlations between uncertainty, not causality. 
Thus, it can help hypothesize about the sources of uncertainty, but they require additional validation and clinical feedback. 
Last but not least, the proposed analysis does not need to be limited to instance segmentation. It can be performed for structure- or patient-wise uncertainty values for any semantic segmentation task, as well as for an image classification.
\bibliographystyle{splncs04}
\bibliography{bibliography}

\begin{thebibliography}{10}
\providecommand{\url}[1]{\texttt{#1}}
\providecommand{\urlprefix}{URL }
\providecommand{\doi}[1]{https://doi.org/#1}

\bibitem{osti_1561669}
Begoli, E., Bhattacharya, T., Kusnezov, D.F.: The need for uncertainty quantification in machine-assisted medical decision making. Nature Machine Intelligence  \textbf{1}(1) (2019)

\bibitem{10.1145/3461702.3462571}
Bhatt, U., Antor\'{a}n, J., Zhang, Y., Liao, Q.V., Sattigeri, P., Fogliato, R., Melan\c{c}on, G., Krishnan, R., Stanley, J., Tickoo, O., Nachman, L., Chunara, R., Srikumar, M., Weller, A., Xiang, A.: Uncertainty as a form of transparency: Measuring, communicating, and using uncertainty. In: Proceedings of the 2021 AAAI/ACM Conference on AI, Ethics, and Society. p. 401–413. AIES '21, Association for Computing Machinery, New York, NY, USA (2021)

\bibitem{uncs_survey}
Gawlikowski, J., Tassi, C., Ali, M., Lee, J., Humt, M., Feng, J., Kruspe, A., Triebel, R., Jung, P., Roscher, R., Shahzad, M., Yang, W., Bamler, R., Zhu, X.: A survey of uncertainty in deep neural networks. Artificial Intelligence Review pp. 1--77 (2023)

\bibitem{grabner-2006}
Grabner, G., Janke, A.L., Budge, M.M., Smith, D., Pruessner, J., Collins, D.L.: Symmetric atlasing and model based segmentation: An application to the hippocampus in older adults. In: Larsen, R., Nielsen, M., Sporring, J. (eds.) Medical Image Computing and Computer-Assisted Intervention -- MICCAI 2006. pp. 58--66. Springer Berlin Heidelberg, Berlin, Heidelberg (2006)

\bibitem{10.3389/fnins.2020.00282}
Jungo, A., Balsiger, F., Reyes, M.: Analyzing the quality and challenges of uncertainty estimations for brain tumor segmentation. Frontiers in Neuroscience  \textbf{14} (2020)

\bibitem{LAROSA2020102335}
La~Rosa, F., Abdulkadir, A., Fartaria, M.J., Rahmanzadeh, R., Lu, P.J., Galbusera, R., Baraković, M., Thiran, J., Granziera, C., Cuadra, M.B.: {Multiple sclerosis cortical and WM lesion segmentation at 3T MRI: a deep learning method based on FLAIR and MP2RAGE}. NeuroImage: Clinical  \textbf{27},  102335 (2020)

\bibitem{DojatMICCAI}
Lambert, B., Forbes, F., Doyle, S., Tucholka, A., Dojat, M.: Beyond voxel prediction uncertainty: Identifying brain lesions you can trust. International Workshop on Interpretability of Machine Intelligence in Medical Image Computing pp. 61--70 (2022)

\bibitem{lambertreview}
Lambert, B., Forbes, F., Tucholka, A., Doyle, S., Dehaene, H., Dojat, M.: {Trustworthy clinical AI solutions: A unified review of uncertainty quantification in Deep Learning models for medical image analysis}. Artificial intelligence in medicine  \textbf{150},  102830 (4 2024)

\bibitem{madsen-2021}
Madsen, M.A., Wiggermann, V., Bramow, S., Christensen, J.R., Sellebjerg, F., Siebner, H.R.: {Imaging cortical multiple sclerosis lesions with ultra-high field MRI}. NeuroImage. Clinical  \textbf{32},  102847 (2021)

\bibitem{shifts20}
Malinin, A., Athanasopoulos, A., Barakovic, M., Cuadra, M.B., Gales, M.J.F., Granziera, C., Graziani, M., Kartashev, N., Kyriakopoulos, K., Lu, P.J., Molchanova, N., Nikitakis, A., Raina, V., Rosa, F.L., Sivena, E., Tsarsitalidis, V., Tsompopoulou, E., Volf, E.: Shifts 2.0: Extending the dataset of real distributional shifts (2022)

\bibitem{molchanova_isbi}
Molchanova, N., Raina, V., Malinin, A., La~Rosa, F., Muller, H., Gales, M., Granziera, C., Graziani, M., Cuadra, M.B.: Novel structural-scale uncertainty measures and error retention curves: Application to multiple sclerosis. In: 2023 IEEE 20th International Symposium on Biomedical Imaging (ISBI). pp.~1--5 (2023)

\bibitem{molchanova_preprint}
Molchanova, N., Raina, V., Malinin, A., Rosa, F.L., Depeursinge, A., Gales, M., Granziera, C., Muller, H., Graziani, M., Cuadra, M.B.: Structural-based uncertainty in deep learning across anatomical scales: Analysis in white matter lesion segmentation (2024)

\bibitem{TAlber}
Nair, T., Precup, D., Arnold, D., Arbel, T.: Exploring uncertainty measures in deep networks for multiple sclerosis lesion detection and segmentation. Medical Image Analysis  \textbf{59},  101557 (2020)

\bibitem{9206659}
Rottmann, M., Colling, P., Paul~Hack, T., Chan, R., Hüger, F., Schlicht, P., Gottschalk, H.: Prediction error meta classification in semantic segmentation: Detection via aggregated dispersion measures of softmax probabilities. In: 2020 International Joint Conference on Neural Networks (IJCNN). pp.~1--9 (2020)

\bibitem{mac}
Thompson, A.J., Banwell, B., Barkhof, F., Carroll, W.M., Coetzee, T., Comi, G., Correale, J., Fazekas, F., Filippi, M., Freedman, M.S., Fujihara, K., Galetta, S., Hartung, H.P., Kappos, L., Lublin, F., Marrie, R.A., Miller, A., Miller, D.H., Montalbán, X., Mowry, E.M., Sørensen, P.S., Tintoré, M., Traboulsee, A., Trojano, M., Uitdehaag, B.M.J., Vukusic, S., Waubant, E., Weinshenker, B.G., Reingold, S.C., Cohen, J.A.: {Diagnosis of multiple sclerosis: 2017 revisions of the McDonald criteria}. Lancet Neurology  \textbf{17}(2),  162--173 (2018)

\end{thebibliography}

\end{document}